\documentclass[aps,showpacs,pra]{revtex4-1}
\usepackage{amsmath}
\usepackage{graphicx}
\usepackage{epstopdf}
\usepackage{amsfonts}
\usepackage{amssymb}

\begin{document}
\bibliographystyle{unsrt}

\title{Singlet States Preparation for Three $\Lambda$-type Atoms with Rydberg Blockade Mechanism}
\author{Rong-Can Yang$^{1,2}$}
\email{rcyang@fjnu.edu.cn}
\author{Xiu Lin$^{1,2}$}
\author{Hong-Yu Liu$^{3}$}
\affiliation{$^1$ Fujian Provincial Key Laboratory of Quantum Manipulation and New Energy Materials,\\ College of Physics and Energy, Fujian Normal University, Fuzhou, 350007, China}
\affiliation{$^2$ Fujian Provincial Collaborative Innovation Center for Optoelectronic Semiconductors and Efficient Devices, Xiamen 361005, China}
\affiliation{$^3$ College of Science, Yanbian University, Yanji, 133002, China}
\date{\today}

\begin{abstract}
A proposal for the generation of singlet states of three $\Lambda$-type Rydberg atoms is presented.
The singlet state is prepared through the combination of a Rydberg state and an EPR pair, and
the scheme relies on the Rydberg blockade effect which prevents the simultaneous excitation of the two atoms to a Rydberg state.
In addition, some frequency detuning between lasers and atomic transitions is set to eliminate the degenerate of the two ground states.
And finally, a series of numerical simulations are made to show the feasibility of the scheme.
\end{abstract}


\maketitle

\section{Introduction}
Entanglement is a particular feature for quantum physics. It can be used not only to test violations of locality, but also for the implementation of quantum information processing \cite{Einstein}, quantum computation \cite{Barenco}, quantum measurement \cite{Caracciolo}, and so on.
So far, much progress has been made for entanglement, especially for entangled qubits because qubits' concepts and implementation may be the most simply.
Although it is true, we should not neglect entangled high-dimension states since they have more peculiar properties than entangled qubits. As a special example, supersinglets $\left| {S_N^{(d)}} \right\rangle$, the entangled states of total spin zero of N particles of spin $(d-1)/2$ \cite{Cabelloa, Cabellob}, attract some attention.
These antisymmetric states are N-lateral rotationally invariant.
Some types of them are shown to be in connection with violations of Bell's inequalities \cite{Mermin}, some of them are used in two-particle proofs of Bell¡¯s theorem without inequalities \cite{Cabelloc}, and so on.
Amongst them, N-particle N-level singlet states $|S_N^{(N)}\rangle$ are particularly fascinating, which can be expressed as  \cite{Cabelloa}
\begin{equation}\label{sn}
|S_N^{(N)}\rangle = \frac{1}{{\sqrt {N!} }}\sum\limits_{\scriptstyle permutations\hfill\atop
\scriptstyle of\,01 \cdots \left( {N - 1} \right)\hfill} {{{\left( { - 1} \right)}^t}\left| {ij \cdots n} \right\rangle },
\end{equation}
where $t$ is the number of transpositions of pairs of elements that must be composed to place the elements in canonical order (i.e. $0, 1,2, \cdots, N-1 $).
The distinctive application of these states was to solve ``N-strangers", ``secret sharing" and ``liar detection" problems which have no solution using classical tools \cite{Cabelloa, Cabellob}.
In addition, they can be applied to realize an unknown unitary transformation \cite{Weinfurter} and construct decoherence-free subspaces \cite{Cabellob,Zanardi}, etc.

For the preparation of singlet states, as Cabello remarked \cite{Cabelloa, Cabellob}, in spit of lots of potential applications of $|S_N^{(N)}\rangle$, it was a formidable physical challenge to prepare these states for $N \geqslant 3$. Although there had still been some proposals presented for the generation of $|S_3^{(3)}\rangle$ with cavity QED \cite{JinGS, LinGW} and ion-trapped technologies \cite{HuangXH}, most of which were complex because several steps are needed to complete the whole procedure and hard to be generalized for $|S_N^{(N)}\rangle$.
However, if Eq.(\ref{sn}) is rechecked, then we will find $|S_2^{(2)}\rangle= \frac{1}{\sqrt{2}}(|01\rangle-|10\rangle)$ which is a simplest entangled state (EPR pair).
So, whether can we generate $|S_3^{(3)}\rangle$ on the basis of $|S_2^{(2)}\rangle$? In 2009, we first put forward an ``addition strategy" \cite{YangRC1}, i.e. a separated particle simultaneously interacts with two ones initially in the EPR pair.
Since then, a series of schemes have successively been proposed \cite{Shao,YangRCQINP1, ZhangSAP, xiaYSR, xiaYJMO}.

Rydberg blockade mechanism, an effect that can prevent the simultaneous flow or excitation of the atoms by shifting the double-excitation states due to the long range dipole-dipole interaction characteristics of Rydberg atoms, was first pointed out by the Zoller group as a crucial ingredient for quantum gates, either using single neutral atoms \cite{Jaksch} or macroscopic samples \cite{Lukin}.
Now, this mechanism has become a focus in the field of quantum information science \cite{Comparat,Saffman,Low}, including providing a generic mechanism for the control of quantum states including entanglement of two or more particles and the realization of quantum gates.
In earlier 2016, we suggested to generate singlet states for $\Xi$-type Rydberg atoms \cite{Yang1}.
Whether can we utilize $\Lambda$-type ones to obtain $S_3^{(3)}$?
To achieve this aim, most of time, we should eliminate the degenerate situation of the ground states.
In this paper, we will show that we can use appropriate frequency detuning to eliminate degeneration and generate three $\Lambda$-type $S_3^{(3)}$ with adiabatic passage.
Our method also provide a novel idea for other entangled states preparation for $\Lambda-$type atoms with Rydberg blockade mechanism.

The paper is organized as follows.
In Sec. II, we describe the details for preparing $S_3^{(3)}$.
Then numerous simulation is shown and analyzed in Sec. III.
Finally in Sec. IV, the brief feasibility is discussed and a conclusion is made.
\section{Generation of three-qutrit singlet states}
\begin{figure}
  \centering
  \includegraphics[width=8cm]{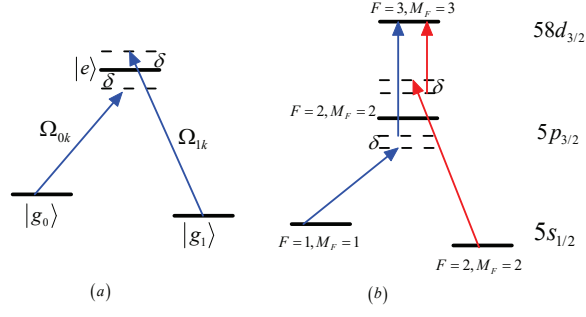}\\
  \caption{(Color online) Atomic level configurations and transitions, in which (a) Direct transition and (b) two-photon transition between two ground states and an excited states}\label{lct}

\end{figure}

Suppose that three $\Lambda$-type Rydberg atoms are trapped in three separate optical potentials with the distance being so close that the dipole-dipole interaction for each two atoms exists.
As is shown in Fig.\ref{lct}a, each atom has a Rydberg level $\left|{e}\right\rangle$ and two ground levels $\left|{g_0}\right\rangle$ and $\left|{g_1}\right\rangle$.
For the sake of clarity, we mark each atom with the subscript $k \left({k=1,2,3}\right)$.
The transition for each atom is driven by two non-resonant laser pulses: one for the transition ${\left|{g_0}\right\rangle}_k \to {\left|{e}\right\rangle}_k$ with time-dependent Rabi frequency ${\Omega}_{0k}\left({t}\right)$ and time-independent dutuning $-\delta$, the other for ${\left|{g_1}\right\rangle}_k \to {\left|{e}\right\rangle}_{k}$ with time-dependent Rabi frequency ${\Omega}_{1k}\left({t}\right)$ and opposite detuning $\delta$.
Under the resolved-sideband limit and rotating-wave approximation, the system Hamiltonian in the rotating picture with respect to $H_0 =
{\hbar} \sum\limits_{k=1}^3 {\omega_e {\left|{e}\right\rangle}_k \left\langle{e}\right| + \left(\omega_{g_0} - \delta \right) {\left|{g_0}\right\rangle}_k \left\langle{g_0}\right| + \left(\omega_{g_1} + \delta \right) {\left|{g_1}\right\rangle}_k \left\langle{g_1}\right|}$ reads $\left(\hbar = 1\right)$
\begin{equation}\label{f1}
\begin{split}
H &= \sum\limits_{j,k = 1;j \ne k}^3 {{U_{jk}}{\left| e \right\rangle}_k \langle e|}  + \delta \sum\limits_{k = 1}^3 {\left( {{{\left| {{g_0}} \right\rangle }_k}\langle {g_0}| - {{\left| {{g_1}} \right\rangle }_k}\langle {g_1}|} \right)} \\
{\rm{     }} &+ \left( {\sum\limits_{k = 1}^3 {{\Omega _{0k}}\left( t \right){{\left| {{g_0}} \right\rangle }_k}\langle e| + {\Omega _{1k}}\left( t \right){{\left| {{g_1}} \right\rangle }_k}\langle e|}  + H.c.} \right),
\end{split}
\end{equation}
with $U_{jk}$ representing the energy shift when $j$th and $k$th atoms are both in the Rydberg levels $\left|{e}\right\rangle$ due to the dipole-dipole interaction.

Taking the strong Rydberg blockade limit into account, i.e. ${U_{jk}} >> \delta, {\Omega _{ml}}\left( t \right), (j,k,l = 1,2,3; j \ne k; m = 0,1)$, only one atom can stay in the Rydberg level.
And for the sake of simplicity, we assume that the second and the third atoms are symmetrical, i.e. $\Omega_{m3}\left(t\right)=\Omega_{m2}\left(t\right), (m=0,1)$.
Therefore, if the system is initially prepared in the partly-entangled state ${\left| e \right\rangle _1}\frac{1}{{\sqrt 2 }}\left( {{{\left| {{g_0}} \right\rangle }_2}{{\left| {{g_1}} \right\rangle }_3} - {{\left| {{g_1}} \right\rangle }_2}{{\left| {{g_0}} \right\rangle }_3}} \right)$, then the system-state evolution will be confined in the subspaces$\left\{ {\left| 1 \right\rangle ,\left| 2 \right\rangle , \cdots ,\left| 7 \right\rangle } \right\}$ without the consideration of any dissipation and decoherence, in which logic states are defined as
\begin{equation}\label{f3}
\begin{array}{l}
\left| 1 \right\rangle  = {\left| {{g_0}} \right\rangle _1}\frac{1}{{\sqrt 2 }}\left( {{{\left| e \right\rangle }_2}{{\left| {{g_1}} \right\rangle }_3} - {{\left| {{g_1}} \right\rangle }_2}{{\left| e \right\rangle }_3}} \right),\\
\left| 2 \right\rangle  = {\left| {{g_0}} \right\rangle _1}\frac{1}{{\sqrt 2 }}\left( {{{\left| {{g_0}} \right\rangle }_2}{{\left| {{g_1}} \right\rangle }_3} - {{\left| {{g_1}} \right\rangle }_2}{{\left| {{g_0}} \right\rangle }_3}} \right),\,\\
\left| 3 \right\rangle  = {\left| e \right\rangle _1}\frac{1}{{\sqrt 2 }}\left( {{{\left| {{g_0}} \right\rangle }_2}{{\left| {{g_1}} \right\rangle }_3} - {{\left| {{g_1}} \right\rangle }_2}{{\left| {{g_0}} \right\rangle }_3}} \right),\\
\left| 4 \right\rangle  = {\left| {{g_0}} \right\rangle _1}\frac{1}{{\sqrt 2 }}\left( {{{\left| {{g_0}} \right\rangle }_2}{{\left| e \right\rangle }_3} - {{\left| e \right\rangle }_2}{{\left| {{g_0}} \right\rangle }_3}} \right),\\
\left| 5 \right\rangle  = {\left| {{g_1}} \right\rangle _1}\frac{1}{{\sqrt 2 }}\left( {{{\left| {{g_0}} \right\rangle }_2}{{\left| {{g_1}} \right\rangle }_3} - {{\left| {{g_1}} \right\rangle }_2}{{\left| {{g_0}} \right\rangle }_3}} \right),\\
\left| 6 \right\rangle  = {\left| {{g_1}} \right\rangle _1}\frac{1}{{\sqrt 2 }}\left( {{{\left| e \right\rangle }_2}{{\left| {{g_1}} \right\rangle }_3} - {{\left| {{g_1}} \right\rangle }_2}{{\left| e \right\rangle }_3}} \right),\,\\
\left| 7 \right\rangle  = {\left| {{g_1}} \right\rangle _1}\frac{1}{{\sqrt 2 }}\left( {{{\left| {{g_0}} \right\rangle }_2}{{\left| e \right\rangle }_3} - {{\left| e \right\rangle }_2}{{\left| {{g_0}} \right\rangle }_3}} \right).
\end{array}
\end{equation}
Accordingly, the Hamiltonian (\ref{f1}) reduces to
\begin{equation}\label{f2}
\begin{split}
  H
   &= \delta {\left|2\right\rangle \left\langle 2 \right|} + 2\delta {\left|4\right\rangle \left\langle 4 \right|}
    - \delta {\left|5\right\rangle \left\langle 5 \right|} - 2\delta {\left|6\right\rangle \left\langle 6 \right|}    \\
   &+ (\Omega_{02}(t){\left|1\right\rangle \left\langle 2 \right|}+ \Omega_{01}(t){\left|2\right\rangle \left\langle 3 \right|}
    + \Omega_{12}(t){\left|2\right\rangle \left\langle 4 \right|}     \\
   &+ \Omega_{11}(t){\left|3\right\rangle \left\langle 5 \right|}+ \Omega_{02}(t){\left|5\right\rangle \left\langle 6 \right|}
    + \Omega_{12}(t){\left|5\right\rangle \left\langle 7 \right|} +H.c.),
\end{split}
\end{equation}

In order to simplify the Hamiltonian described in the equation (\ref{f2}), we transform to the space $\{ \left|1\right\rangle, \left|3\right\rangle, \left|4\right\rangle, \left|6\right\rangle, \left|7\right\rangle$, $ \left|\eta_+\right\rangle, \left|\eta_-\right\rangle \}$ with $\left| {{\eta _ + }} \right\rangle  = \frac{1}{N}\left( {{\Omega _{01}}\left| 2 \right\rangle  + {\Omega _{11}}\left| 5 \right\rangle } \right)$, $\left| {{\eta _ - }} \right\rangle  = \frac{1}{N}\left( {{\Omega _{11}}\left| 2 \right\rangle  - {\Omega _{01}}\left| 5 \right\rangle } \right)$ and \\ $N = \sqrt {\Omega _{01}^2 + \Omega _{11}^2}$, where we have denoted $\Omega_{jk}\left(t\right)$ as $\Omega_{jk}$ for short, as is suitable in the following paragraphs. After that, we will obtain
\begin{equation}\label{f4}
\begin{split}
  H&=\frac{\delta}{N^2}(\Omega_{01}^2-\Omega_{11}^2)(|\eta_+\rangle \langle \eta_+|-|\eta_-\rangle \langle \eta_-|) \\
   &+\frac{2\delta\Omega_{01}\Omega_{11}}{N^2}(|\eta_+\rangle \langle \eta_+|-|\eta_-\rangle \langle \eta_-|)
    +2\delta|4\rangle \langle 4| -2\delta |6 \rangle \langle 6| \\
   &+\left\{\frac{\Omega_{02}}{N}|1\rangle (\Omega_{01}\langle \eta_+|+\Omega_{11}\langle \eta_-|)+N|\eta_+\rangle \langle 3|\right.\\
   &+\frac{\Omega_{12}}{N}(\Omega_{01}|\eta_+\rangle + \Omega_{11} |\eta_-\rangle)\langle 4 |
    +\frac{\Omega_{02}}{N}(\Omega_{11}|\eta_+\rangle - \Omega_{01} |\eta_-\rangle)\langle 6 |\\
   &\left.+\frac{\Omega_{12}}{N}(\Omega_{11}|\eta_+\rangle - \Omega_{01} |\eta_-\rangle)\langle 7 | +H.c.\right\}
\end{split}
\end{equation}

Taking into account of the large-detuning condition $\delta>>\Omega_{0k},\Omega_{1k}$, we can adiabatically neglect the logic states $\left|\eta_+\right\rangle, \left|\eta_-\right\rangle, \left|4\right\rangle, \left|6\right\rangle$. In fact, it should be noted that the large-detuning condition should not be strictly satisfied in our proposal because of selective-transition rules.
If the Hamiltonian is carefully checked, then we can see $\left|\eta_+\right\rangle$ may be populated more likely than that of the other logic excited states. Thus, we can only consider the system evolution in the subspace $\{\left|1\right\rangle, \left|3\right\rangle, \left|7\right\rangle, \left|\eta_+\right\rangle \}$, and obtain
\begin{equation}\label{f5}
\begin{split}
H_{eff} &\approx \frac{\delta}{N^2}(\Omega _{01}^2 - \Omega _{11}^2)|\eta_+\rangle \langle \eta_+|
         + \left(\frac{\Omega_{02}\Omega_{01}}{N}|1\rangle \langle \eta_+|\right.\\
        & \left.+ N |\eta_+ \rangle \langle 3|
                +\frac{\Omega_{11}\Omega_{02}}{N}|\eta_+\rangle \langle 3| + H.c. \right)
\end{split}
\end{equation}

If we further consider the situation of ${\Omega _{1k}} =  {\Omega _{0k}} \left( {k = 1,2} \right)$ , then we can easily obtain a dark state for the effective Hamiltonian, i.e.
 \begin{equation}\label{f6}
 \left| {D\left( t \right)} \right\rangle = \cos{\theta} \left| 3 \right\rangle  - \sin{\theta} \left( \left| 1 \right\rangle  + \left| 7 \right\rangle \right)/\sqrt{2}.
 \end{equation}
with $\tan{\theta}=\sqrt{2}\Omega_{01} / \Omega_{02}$.

With the adiabatic-passage condition $\left| {\dot \theta } \right| << \sqrt {2\Omega _{01}^2 + \Omega _{02}^2} $ satisfied, pulse shapes are designed such that
\begin{equation}\label{f7}
\mathop {\lim }\limits_{t \to  - \infty } \frac{{{\Omega _{01}}\left( t \right)}}{{{\Omega _{02}}\left( t \right)}} = 0, \mathop {\lim }\limits_{t \to  - \infty } \frac{{{\Omega _{01}}\left( t \right)}}{{{\Omega _{02}}\left( t \right)}} = 1,
\end{equation}
then the system state can be adiabatically transferred from the initial state $\left|3\right\rangle$ to the three-qutrit singlet state
\begin{equation}\label{f8}
\begin{split}
\left|S_3\right\rangle
  &= \frac{1}{\sqrt{3}} \left(\left|3\right\rangle - \left|1\right\rangle - \left|7\right\rangle\right)\\
  &=\frac{1}{\sqrt{6}}\left(\left|{g_0g_1e}\right\rangle - \left|{g_1g_0e}\right\rangle +\left|{g_1eg_0}\right\rangle \right.\\
  &\left.- \left|{g_1g_0e}\right\rangle +\left|{eg_0g_1}\right\rangle - \left|{eg_1g_0}\right\rangle\right)_{123}
\end{split}
\end{equation}

\section{Numerical simulation and analysis}
In this section, let us simulate the system-state evolution and make some analysis. To achieve this aim, we add two two-excitations vectors $|8\rangle = {|e\rangle}_1 \frac {1}{\sqrt{2}}{(|{g_0e}\rangle-|eg_0\rangle)}_{23}, |9\rangle = {|e\rangle}_1 \frac {1}{\sqrt{2}}{(|{g_1e}\rangle-|eg_1\rangle)}_{23}$ to Eq. (\ref{f3}), and then change the Hamiltonian described in Eq. (\ref{f2}) to
\begin{equation}\label{f10}
  \begin{split}
    H' &= H +(U+\delta) |8\rangle \langle 8| + (U-\delta) |9\rangle \langle 9| +(-\Omega_{01} |1\rangle \langle 9| \\
       &+ \Omega_{02} |3\rangle \langle 8|- \Omega_{12} |3\rangle \langle 9|+ \Omega_{01} |4\rangle \langle 8|- \Omega_{11} |6\rangle \langle 9| \\
       &+ \Omega_{11} |7\rangle \langle 8| + H.c.).
  \end{split}
\end{equation}
Here we have simply set $U_{12}=U_{13}=U, \Omega_{11} = \Omega_{01},$ and $\Omega_{12} = \Omega_{02}$. Similar to previous schemes, laser pulses used in our proposal are also designed by one Gauss beam or the combination of two Gauss beams with an appropriate choice of parameters \cite{Vitanov}
\begin{equation}\label{f9}
  \begin{split}
    \Omega_{01}\left(t\right) &= \Omega_0 \exp\left({\frac{-\left(t-\tau\right)^2}{T^2}}\right),\\
    \Omega_{02}\left(t\right) &= \Omega_0 \exp\left({\frac{-\left(t-\tau\right)^2}{T^2}}\right) + \Omega_0 \exp\left({\frac{-\left(t+\tau\right)^2}{T^2}}\right),
  \end{split}
\end{equation}
where $\Omega_0$ represents the time-independent amplitude and $T$ ($\tau$) the pulse width (delay).

\begin{figure}
  \centering
  \includegraphics[width=8cm]{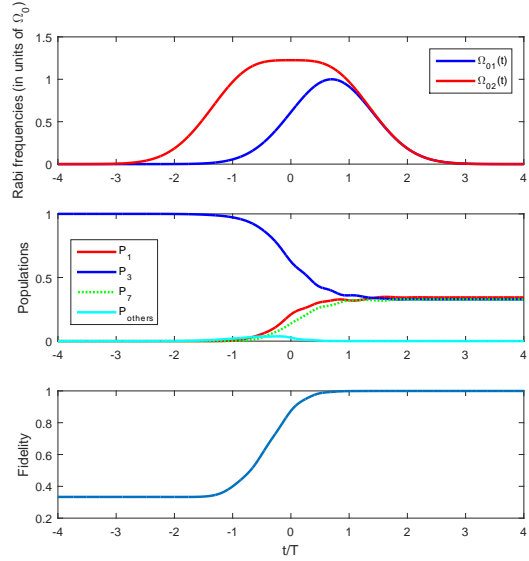}\\
  \caption{(Color online) An explicit example to numerically simulate the generation of $|S_3^{(3)}\rangle$ in ideal condition, where time-dependence Rabi frequencies are shown on the top, the populations for logic states in the middle, and the fidelity on the below.}\label{population}

\end{figure}

To begin with, the system-state evolution under the Hamiltonian described in Eq. (\ref{f10}) is demonstrated in Fig.\ref{population}, where we have chosen $\Omega_0 T = 10, \tau= 0.7 T,U=10\Omega_0$ and $\delta = \Omega_0,$ and expressed the whole system state $\left|\psi\left(t\right)\right\rangle$ at any time as $\left|\psi\left(t\right)\right\rangle = \sum\limits_{k = 1}^9{c_k}\left(t\right)\left|k\right\rangle$ with $\sum\limits_{k = 1}^9{c_k}\left(t\right) = 1$. Then the population for each logic state $|k\rangle$ and the fidelity $F$ for the generation of singlet states are defined as $P_k=|c_k|^2$ and  $F=|\langle S_3|\psi (t)\rangle|^2$, respectively. From Fig.\ref{population}, it is clearly demonstrated that populations of $|1\rangle,|3\rangle,|7\rangle$ will eventually be equal, while the other logic states are virtually excited during the whole procedure. The figure on the bottom further proves that a singlet state for these three Rydberg atoms with Fidelity $99.96\%$ is generated as long as the interaction $t$ is long enough.

\begin{figure}
  \centering
  \includegraphics[width=8cm]{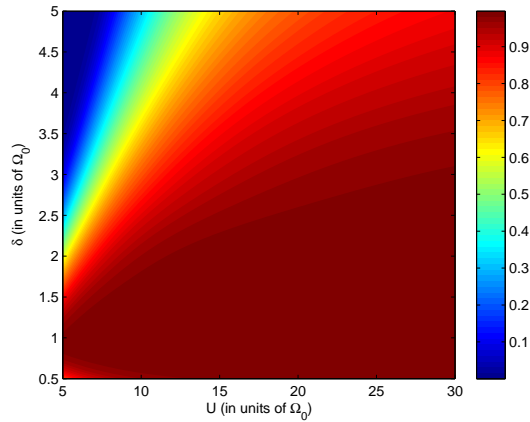}\\
  \caption{(Color online) Influence of two-Rydberg-states energy shift $U$ and detuning $\delta$ on the fidelity.}\label{FUd}
\end{figure}

In addition, the effect of the dipole-dipole interaction $U$ and the detuining $\delta$ on Fidelity $F$ is analyzed, as is shown in Fig.\ref{FUd} with $\Omega_0 T = 10, \tau= 0.7 T,\delta = \Omega_0$ and $t=4T$. From the illustration, it is clearly seen that the higher $U$, the wider range of $\delta$ can be allowed for the generation of singlet states. For example, if $F\geqslant0.98$, then $0.79\Omega_0\leqslant \delta \leqslant 1.08\Omega_0$ should be required when $U=5\Omega_0$, while the range of $\delta$ can be larger with the increase of $\Omega_0$.

\begin{figure}
  \centering
  \includegraphics[width=8cm]{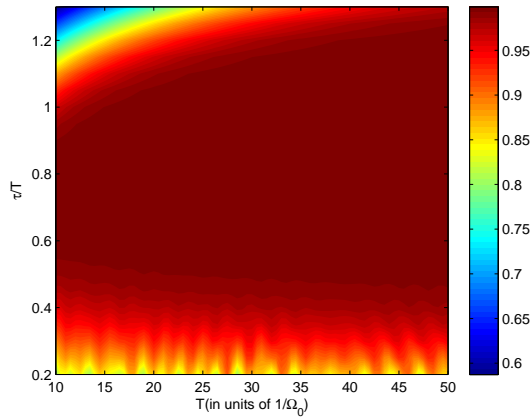}\\
  \caption{(Color online) Fidelity as the function of pulse width $T$ and delay $\tau$.}\label{FTt}
\end{figure}

Moreover, how the shapes of laser pulses influence the Fidelity is also illustrated in Fig. \ref{FTt} with $U=10\Omega_0, \delta = \Omega_0$ and $t=4T$. Obviously, the lower limit of $\tau/T$ is about $0.35$ to obtain $F=0.93$ and $0.5$ for $F=0.99$, while the upper limit of $\tau/T$ to gain the same Fidelity is almost linearly proportional to $\Omega_0T$, and the relation is about $\tau/T < 0.01T + 0.8$.

\begin{figure}
  \centering
  \includegraphics[width=8cm]{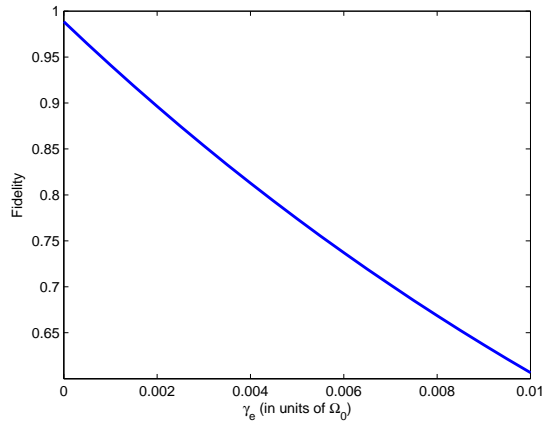}\\
  \caption{(Color online) Effect of spontaneous emission on fidelity.}\label{Fgamma}
\end{figure}

Last but not the least, the effect of the spontaneous emission of Rydberg atoms on the Fidelity $F$ is discussed. To this motivation, we resort to the master equation for density matrix $\rho(t)$ of the whole system, i.e.
\begin{equation}\label{masterequation}
    \begin{split}
        \dot\rho &= - i \left[H, \rho\right] \\
        &- \sum\limits_{p = {g_0},{g_1}} {\sum\limits_{q = 1}^3 {\frac{\gamma }{2}\left( {S_{pq}^ + {S_{pq}}\rho  - 2{S_{pq}}\rho S_{pq}^ +  + S_{pq}^ + {S_{pq}}\rho } \right)} },
    \end{split}
\end{equation}
where $S_{pq}^+ = {|e\rangle}_q \langle p|$, and the decay rate $\gamma$ is assumed to be the same and equal to $\gamma_e/2$ with $\gamma_e$ represents the spontaneous emission rate. The function of fidelity versus $\gamma_e/\Omega_0$ is demonstrated in Fig. \ref{Fgamma}, where we have chosen $\Omega_0 T = 10, \tau= 0.7 T,U=5\Omega_0, \delta = \Omega_0,$ and $t=3T$. From Fig.\ref{Fgamma}, it is illustrated that the Fidelity of the generated singlet states is strongly influenced by atomic spontaneous emission rate $\gamma_e$. For example, the Fidelity $F \approx 0.96 $ with the chosen of $\gamma_e = 0.001 \Omega_0$, while it quickly decreases to $F \approx 0.61$ when $\gamma_e = 0.01 \Omega_0$. So the scheme may be feasible only if the spontaneous emission rate $\gamma_e$ is much smaller than $\Omega_0$. Luckily, it can be realized because the Rydberg levels has a long lifespan, which will further discussed in next section.

\section{Feasibility and Conclusion}
We now make a short discussion on the experimental feasibility.
The three levels $|g_0\rangle$, $|g_1\rangle$, and $|e\rangle$ for Rydberg atoms can be represented by two long-lived sublevels $|5s_{1/2}, F=1, M_F=1\rangle$, $|5s_{1/2}, F=2, M_F=2\rangle$ and a Rydberg level $|58d_{3/2},F=3, M_F=3\rangle$ of $^{87}Rb$.
The transition $|g_0\rangle \left(|g_1\rangle\right) \to |e\rangle$ can be achieved by a nearly two-photon transition with a $\sigma^+ (\pi)$-polarized laser at 795nm and two $\sigma^+$-polarized lasers at 475nm. Please see the Fig.\ref{lct}b.
The frequency of the 795nm laser is largely red(blue)-detuned by $\Delta_1-\delta (\Delta_2+\delta)$ with $\Delta_{1(2)}>>\delta>0$ from the transition from $|g_0\rangle \left(|g_1\rangle\right)$ to the excited state $|5p_{3/2}, F=2, M_F=2\rangle$ in order to reduce spontaneous emission, similar to Ref.\cite{Wilk2010}.
According to Ref.\cite{Wilk2010,Gaetan2009}, when each two of the three optical tweezers are separated by 4$\mu m$, energy shift due to Rydberg blockade mechanism is about $50MHz$ and the measured Rabi frequency of the two-photon transition from $|g_{0,1}\rangle \to |e\rangle$ is about $\Omega_0=6MHz$.
If we also choose $\delta$ to be $6MHz$, then the required time is about $10\mu s$ and the fidelity of the generated singlet state is about $0.90$ with the consideration of the life of Rydberg level induced by blackbody radiation $(\approx 160 \mu s)$ and the radiative decay time $(\approx 200 \mu s)$ \cite{Wilk2010, Gallagher1994}. Therefore, our scheme may be feasible based on current technologies.

In conclusion, a singlet state for three $\Lambda$-type atoms is prepared with Rydberg blockade mechanism.
The singlet state is generated by the combination of a Rydberg atom and an EPR pair, which may be scalable.
In the present proposal, some detuning between lasers and atomic transitions is used to eliminate the degenerate of ground states.
Compared with proposals with cavity QED or trapped ions, the present one doesn't need any media to store quantum information.
And finally, numerical simulation is made, showing our proposal may be feasible.

\section{Acknowledgments}
   This work is supported by the National Natural Science Foundation of China (Grant nos. 61308012 and 61275215).



\begin{thebibliography}{99}
\bibitem{Einstein}A. Einstein, B. Podolsky, and N. Rosen, Phys. Rev. 47 (1935) 777.
\bibitem{Barenco}A. Barenco, C.H. Bennett, R. Cleve, D.P. DiVincenzo, N. Margolus, P. Shor, T. Sleator, J.A. Smolin, and H. Weinfurter, Phys. Rev. A 52 (1995) 3457.
\bibitem{Caracciolo}S. Caracciolo, and A. Pelissetto, Phys. Lett. B 207 (1988) 468
\bibitem{Cabelloa}A. Cabello, Phys. Rev. Lett. 89 (2002) 100402.
\bibitem{Cabellob}A. Cabello, J. Mod. Opt. 50 (2003) 1049.
\bibitem{Mermin}N.D. Mermin, Phys. Rev. D 22 (1980) 356.
\bibitem{Cabelloc}A. Cabello, Phys. Rev. A 58 (1998) 1687.
\bibitem{Weinfurter} H. Weinfurter, and M. \ifmmode \dot{Z}\else \.{Z}\fi{}ukowski, Phys. Rev. A 64 (2001) 010102.
\bibitem{Zanardi}P. Zanardi, and M. Rasetti, Phys. Rev. Lett. 79 (1997) 3306
\bibitem{JinGS}G.S. Jin, S.S. Li, S.L. Feng, and H.Z. Zheng, Phys. Rev. A 71 (2005) 034307.
\bibitem{LinGW}G.W. Lin, M.Y. Ye, L.B. Chen, Q.H. Du, and X.M. Lin, Phys. Rev. A 76 (2007) 014308.
\bibitem{HuangXH}X.H. Huang, G.W. Lin, M.Y. Ye, Y.X. Tang, and X.M. Lin, Opt. Commun. 281 (2008) 4545.
\bibitem{YangRC1}R.C. Yang, X. Lin, Z.P. Huang, and H.C. Li, Opt. Commun. 282 (2009) 1952.
\bibitem{Shao}X.Q. Shao, H.F. Wang, L. Chen, S. Zhang, Y.F. Zhao, and R.H. Yeon, New J. Phys. 12 (2010) 65.
\bibitem{YangRCQINP1}R.C. Yang, L.X. Ye, X. Lin, X. Chen, and H.Y. Liu, Quantum Inf. Process. 14 (2015) 4449.
\bibitem{ZhangSAP}D.Y. Wang, J.J. Wen, S. Hu, W.X. Cui, H.F. Wang, and A.D. Zhu, and S. Zhang, Ann. Phys. 360 (2015) 228.
\bibitem{xiaYSR}Z. Chen,  Y.H. Chen, Y. Xia, J. Song, and B.H. Huang, Sci. Rep. 6 (2016) 22202.
\bibitem{xiaYJMO}Z. Chen, Y.H. Chen, and Y. Xia, J. Mod. Opt. 63 (2016) 92.
\bibitem{Jaksch}D. Jaksch, J.I. Cirac, P. Zoller, S.L. Rolston, R. C\^ot\'e, and M.D. Lukin, M. D, Phys. Rev. Lett. 85 (2000) 2208.
\bibitem{Lukin}M.D. Lukin, M. Fleischhauer,R. Cote, L.M. Duan, D. Jaksch, J.I. Cirac, and P. Zoller, Phys. Rev. Lett. 87 (2001) 037901.
\bibitem{Comparat}D. Comparat, and P. Pillet, J. Opt. Soc. Am. B 27 (2010) A208.
\bibitem{Saffman}M. Saffman, T.G. Walker, and K. M\o{}lmer, Rev. Mod. Phys. 82 (2010) 2313.
\bibitem{Low}R. L{\"o}w, H. Weimer, J. Nipper, J.B. Balewski, B. Butscher, H.P. B{\"u}chler, and T. Pfau, J. Phys. B:  At. Mol. Opt. Phys. 45 (2012) 113001.
\bibitem{Yang1}R.C. Yang, X. Lin, L.X. Ye, X. Chen, J. He, and H.Y. Liu, Quantum Inf. Process. 15(2016) 731.
\bibitem{Vitanov}N.V. Vitanov, K.A. Suomien, and B.W. Shore, J. Phys. B: At. Mol. Opt. Phys. 32 (1999) 4535.
\bibitem{Wilk2010}T. Wilk, A. Ga\"etan, C. Evellin, J. Wolters, Y. Miroshnychenko, P. Grangier, A. Browaeys, Phys. Rev. Lett. 104 (2010) 010502.
\bibitem{Gaetan2009}A. Ga{\"e}tan, Y. Miroshnychenko, T. Wilk, A. Chotia, M. Viteau, D. Comparat, P. Pillet, A. Browaeys, and P. Grangier, Nat. Phys. 5(2009) 115.
\bibitem{Gallagher1994}T.F. Gallagher, Rydberg Atoms, Cambridge University Press, Cambridge, 1994.

\end{thebibliography}
\end{document}